\documentclass[reprint,showkeys,showpacs,preprintnumbers,amsmath,amssymb,
 aps]{revtex4-1}

\usepackage{graphicx}
\usepackage{dcolumn}
\usepackage{bm}
\usepackage{hyperref}
\usepackage{xcolor}

\begin{document}

\title{Anisotropic  n-dimensional quantum cosmological model with fluid}
\author{Sachin Pandey}
\email{sp13ip016@iiserkol.ac.in}
\affiliation{Department of Physical Sciences, \\
Indian Institute of Science Education and Research Kolkata, \\
Mohanpur, West Bengal 741246, India.}

\begin{abstract}
In the present work, we discuss the Wheeler-DeWitt quantization scheme for an n-dimensional anisotropic cosmological model with a perfect fluid in presence of a massless scalar field. We identify the time parameter using the generalization of Schutz formalism and find the wavepacket of the universe following standard Wheeler-DeWitt methodology of quantum cosmology. We also calculate the expectation values of scale factors and volume element. We show that quantized model escapes the singularity and supports bouncing universe solutions.
\end{abstract}
\vspace{2pc}
\maketitle

\section{Introduction}
Higher dimensional models had been investigated widely in the past in order to find a theory to unify gravity with other fundamental forces of physics. It started with Kaluza and Klein's assertion that a fifth dimensions in general relativity will unify gravity with the electromagnetic field\cite{kalu,klein,klein1}. Further motivation of resrting to higher dimensional models came from the expectation of unifying gravity with non-Abelian gauge fields\cite{o1,o2}. Later, spacetime having more than four dimesnions were motivated by 10-dimensional superstring theory and 11-dimensional supergravity theory\cite{ss,sg}.

Quantum cosmology has provided a way to study the early phase of the universe. In the absence of a well accepted quantum theory of gravity, it is quite imperative to look at the quantum behaviour in individual gravitational systems. Wheeler-DeWitt equation is a widely used framework where standard quantum mechanics is used in cosmological models\cite{witt,wheeler,misner}. This framework of quantization suffers from various conceptual problems, which are discussed quite elaborately in the literature\cite{wilt, jj,pinto}. The problem of non-unitary evolution and the absence of properly oriented time are amongst them. The latter is dealt with in various ways\cite{nu1,nu2,nu3,nu4,nu5,nu6}. One of the ways of finding a nice time parameter is to use the evolution of the cosmic fluid following Schutz formalism\cite{schutz1,schutz2}. This method has been widely used as time parameter in quantization of cosmological models\cite{is1,is2,is3,vakili1}. In fact, the alleged loss of unitarity in anisotropic quantum cosmological models has been quite effectively resolved\cite{sp1,sp2,sp3,sp5,sachin} using Schutz formalism. In cosmological models related to Brans-Dicke theory\cite{brans,dicke}, the evolution of the scalar field have been used as time parameter in order to quantize the same without adding any additional matter\cite{vakili2,sachin1,sachin2}. The Brans-Dicke theory of gravity has also been quantized using Schutzâ€™s formalism\cite{bdfluid,sp4}.   
So long as a method gives rise to an oriented scalar time parameter, it is quite an effective method.

In a very recent work, Alves-Junior {\it et al}\cite{ndaniso} have shown the canonical quantization of n-dimensional anisotropic model coupled with a massless scalar field in the absence of any fluid where they have emphasized on natural identification of scalar field as time parameter for the evolution of quantum variables.\\

In the present work, we try to investigate the quantum cosmological solutions of an n-dimensional anisotropic model in which massless scalar field is minimally coupled with gravity in presence of a barotropic perfect fluid($P=\alpha \rho$). The idea is to generalize the work of Alves-Junior {\it et al} to include a fluid, as well as to generalize the recent work on anisotropic cosmologies by Pal and Banerjee\cite{sp1, sp2} to higher dimensions. Keeping in mind the importance of various aspects of higher dimensions, it is quite relevant to ask the questions like whether singularity-free n-dimensional quantum cosmological models realized. Also, this investigation should find if the general result, given by Pal and Banerjee\cite{sp5}, that it is always possible to find a self-adjoint extension of homogeneous quantum cosmologies in four dimensions.\\

In section II,  we write Lagrangian for the gravity sector coupled with a scalar field, followed by the mention of Schutz formalism to n-dimensional models. Similar kind of generalization of Schutz fluid was done by Letelier and Pitelli\cite{ndfrw} in the context of the quantization of an n-dimensional FLRW model. We construct the Hamiltonian using canonical transformation in subsection B of this section. In section III,  canonical quantization of the model is presented for a stiff fluid($\alpha=1$). We obtain the wave packet and expectation values of scale factor and volume element. Further, we are also able to show the unitary evolution of the model with general fluid($\alpha \neq 1$). This is done for an example with $n=7$ which is mathematically tractable. In the end, in section IV, we discuss our results and make some remarks.

\newpage

\section{An n-dimensional cosmological model}
\subsection{Lagrangian of the model}
We start with the Hilbert-Einstein action in n-dimension with a minimally coupled scalar field in the presence of a fluid,
\begin{equation}\label{action}
A=\int d^nx \sqrt{-g}(R+\omega g^{\mu\nu}\phi_{,\mu}\phi_{,\nu})+\int d^nx \sqrt{-g}P, 
\end{equation}
where $\omega$ is a dimensionless parameter and $\phi$ is the massless scalar field. The last term of equation (\ref{action}) represents the matter contribution coming from fluid pressure $P$ related with density $\rho$ by the equation of state $P=\alpha \rho$.
The metric for an n dimensional ($n>4$) spatially homogeneous but anisotropic cosmology is chosen as  
\begin{equation}\label{met}
ds^2=N(t)^2dt^2-a(t)^2(dx^2+dy^2+dz^2)-b(t)^2\sum^{n-4}_{i=1} dl_i^2,
\end{equation}
where $N(t)$ is the lapse function, $a(t)$ is our good old usual scale factor used in flat FLRW cosmology and $b(t)$ is scale factor coming from remaining $(n-4)$ dimensions. So the usual 3-space is isotropic, but the extra dimensions, though isotropic in itself, but anisotropic from the usual 3-space section. \\
The action for scalar gravity part can be written in a reduced form for metric given by (\ref{met}) as 
\begin{multline} \label{gaction}
A_g=V_0\int dt \frac{6}{N}\biggl[ -\dot{a}^2ab^{n-4}-(n-4)\dot{b}\dot{a}a^2b^{n-5}\\-\frac{(n-4)(n-5)}{6}\dot{b}^2a^3b^{n-6}  +\frac{\omega}{6}a^3b^{n-4}\dot{\phi}^2\biggr],
\end{multline} 
where an overdot represents a derivative with respect to coordinate $t$ and $V_0$ denotes $(n-1)$ dimensional volume. We have ignored the surface terms as they do not contribute to the field equations.\\
From (\ref{gaction}), we can write the Lagrangian for the gravity sector as, 
\begin{multline} \label{lg}
L_g= -\frac{6}{N}\dot{a}^2ab^{n-4}-\frac{6(n-4)}{N}\dot{b}\dot{a}a^2b^{n-5}\\-\frac{(n-4)(n-5)}{N}\dot{b}^2a^3b^{n-6}+\frac{\omega}{N}a^3b^{n-4}\dot{\phi}^2.
\end{multline}

Schutz \cite{schutz1, schutz2} showed that four dimensional velocity vector of a perfect fluid can be written using six thermodynamic quantities as,
\begin{equation} \label{4v}
U_\nu=\frac{1}{\mu}\left(\epsilon_{,\nu}+\zeta \beta_{,\nu}+\theta S_{,\nu} \right),
\end{equation} 
where $\mu$ is specific enthalpy, $S$ is specific entropy, $\zeta$ and $\beta$ are potentials connected with rotation and $\epsilon$ and $\theta$ are potentials with no clear physical meaning. For n dimensional velcoity vector, we can consider additional potential terms\cite{ndfrw} like $\zeta$ and $\beta$  in equaution(\ref{4v}) as
\begin{equation} \label{nv}
U_\nu=\frac{1}{\mu}\left(\epsilon_{,\nu}+\zeta \beta_{,\nu}+\theta S_{,\nu} +\sum^{n-4}_{i=1}\zeta_i \beta_{i,\nu}\right).
\end{equation}  Since the potentials $\zeta$, $\zeta_i$, $\beta$ and$\beta_i$  are related to rotation, we can neglect them in the present case as there is no vorticity in the spacetime given by equation  (\ref{met}). Following the widely used technique, developed by Lapchinskii and Rubakov \cite{Lapchinskii} by writing the fluid pressure in terms of potentials given in equation (\ref{nv}), the fluid sector action given in equation (\ref{action}) can be written as
\begin{multline}
 A_f=\int dt \biggl[ N^{-1/\alpha} a^3 b^{n-4}\frac{\alpha}{(\alpha +1)^{1/\alpha +1}}(\dot{\epsilon}+\theta\dot{S})^{1/\alpha +1} \\ exp(-S/\alpha)\biggr].
 \end{multline}
From above equation, Lagrangian for fluid sector can be written as
 \begin{multline} \label{lf}
 L_f=\int dt \biggl[ N^{-1/\alpha} a^3 b^{n-4}\frac{\alpha}{(\alpha +1)^{1/\alpha +1}}(\dot{\epsilon}+\theta\dot{S})^{1/\alpha +1}\\ exp(-S/\alpha)\biggr].
 \end{multline}
 
\subsection{Hamiltonian of the model}  
Following standard procedure \cite{gold}, Hamiltonian for gravity sector from corresponding Lagrangian (\ref{lg}) can be written as
\begin{equation}\label{hg}
H_g=\frac{N}{(n-2)ab^{n-6}}\biggl[ \frac{(n-5)}{12}\frac{p_a^2}{b^2}+\frac{1}{2(n-4)}\frac{p_b^2}{a^2}-\frac{p_ap_b}{2ab}+\frac{(n-2)}{4\omega a^2b^2}p_\phi^2\biggr].
\end{equation}
For fluid sector, Hamiltonian corresponding to Lagrangian (\ref{lf}) can be written as 
\begin{equation} \label{hf}
H_f=\frac{N}{a^{3\alpha}b^{(n-4)\alpha}}p_T
\end{equation}
where $T$ and $p_T$ are related with following canonical transformation
\begin{equation}
T=p_S e^{-S}p_\epsilon^{-(\alpha +1)}, p_T= p_\epsilon^{\alpha +1}e^S, \bar{\epsilon}=\epsilon-(\alpha +1)\frac{p_s}{p_\epsilon}, \bar{p_\epsilon}=p_\epsilon.
\end{equation}
Using (\ref{hg}) and (\ref{hf}), the net Hamiltonian can be written as 

\begin{multline}\label{neth}
\mathcal{H} = \frac{N}{a^3b^{(n-4)}}\biggl[\frac{(n-5)a^2}{12(n-2)}p_a^2+\frac{b^2}{2(n-4)(n-2)}p_b^2 \\-\frac{ab}{2(n-2)}p_ap_b+\frac{1}{4\omega}p_\phi^2+a^{3-3\alpha}b^{(n-4)-(n-4)\alpha}p_T\biggr].
\end{multline}
 The above expression  for $\mathcal{H}$ may look formidable, but can be written in a simpler form with the help of following canonical transformations, 
 \begin{equation}
 A=lna, p_A=ap_a, B=lnb, p_B=bp_b,
\end{equation}
as
\begin{multline} \label{nth}
\mathcal{H} = \bar{N} \biggl[ \frac{(n-5)}{12(n-2)}p_A^2-\frac{1}{2(n-2)}p_A p_B+\frac{1}{2(n-2)(n-4)}p_B^2 \\ +\frac{1}{4\omega}p^2_\phi +e^{(3\alpha-3)A+((n-4)\alpha-(n-4))B}p_T \biggr],
\end{multline}
where $\bar{N}=\frac{N}{a^3b^{n-4}}$.

\section{Quantization of the model}
By varying the action with respect to $\bar{N}$, Hamiltonian constarint can be written as $\hat{H}=\frac{\mathcal{H}}{\bar{N}}=0$. Now the coordinates ($A, B, \phi$ and their corresponding conjugate momenta are promoted to operators using the standard commutation relations ($[q_i, p_i] = i$ in the units $\frac{h}{2\pi} = 1$). $\hat{H}$ can be written in terms of these operators using eq (\ref{nth}) and the Wheeler-DeWitt equation ($\hat{H}\psi=0$) looks like,
\begin{multline}\label{wd}
\biggl[ -\frac{(n-5)}{12(n-2)}\frac{\partial^2}{\partial A^2}  +\frac{1}{2(n-2)} \frac{\partial^2}{\partial A \partial B}\\ -\frac{1}{2(n-2)(n-4)} \frac{\partial^2}{\partial B^2}  -\frac{1}{4\omega}\frac{\partial^2}{\partial \phi^2}\\ -ie^{(3\alpha-3)A+((n-4)\alpha-(n-4))B}\frac{\partial}{\partial T}\biggr]\psi(A,B,\phi, T)=0.
  \end{multline}
For $n=5$, the first term of equation (\ref{wd}) will vanish and Wheeler-DeWitt equation becomes simpler. Hence we shall be considering it later. \\

With ansatz, $\psi(A,B,\phi, T)=\xi(A,B,\phi)e^{-iET}$,  equation (\ref{wd}) takes following form,
\begin{multline} \label{wdw}
\biggl[ \left(2-\sqrt{\frac{6(n-4)}{(n-5)}}\right)\frac{\partial^2}{\partial u^2}+\left(2+\sqrt{\frac{6(n-4)}{(n-5)}}\right)\frac{\partial^2}{\partial v^2}\\+\frac{1}{4\omega}\frac{\partial^2}{\partial \phi^2}+exp\biggl\{ (3\alpha-3)\biggl( \sqrt{\frac{n-5}{12(n-2)}}\frac{u+v}{2}\biggr)\\+((n-4)\alpha-(n-4))\biggl(  \sqrt{\frac{1}{2(n-2)(n-4)}}\frac{u-v}{2}\biggr)\biggr\}\\ E \biggr] \xi(u,v,\phi)=0,
\end{multline}
where $u$ and $v$ are given as,
\begin{eqnarray}
u=\sqrt{\frac{12(n-2)}{(n-5)}}A+\sqrt{2(n-2)(n-4)}B, \\
v=\sqrt{\frac{12(n-2)}{(n-5)}}A-\sqrt{2(n-2)(n-4)}B.
\end{eqnarray}

\subsection{For stiff  fluid $\alpha=1$ and $n>5$} 
For  $\alpha=1$, equation (\ref{wdw}) can be rewritten as
\begin{multline} \label{wdw1}
\biggl[ \left(2-\sqrt{\frac{6(n-4)}{(n-5)}}\right)\frac{\partial^2}{\partial u^2}+\left(2+\sqrt{\frac{6(n-4)}{(n-5)}}\right)\frac{\partial^2}{\partial v^2}\\+\frac{1}{4\omega}\frac{\partial^2}{\partial \phi^2}+E \biggr] \xi(u,v,\phi)=0.
\end{multline}
Solution for equation (\ref{wdw1}) can be written as 
\begin{multline}
\psi_{\lambda,E,k}(u,v,\phi,T)=K \sin(u\sqrt{\frac{\lambda}{|2-\bar{n}|}})\\ \times \sin(v\sqrt{\frac{E-k-\lambda}{2+\bar{n}}})  \sin(\phi\sqrt{4\omega k})e^{-i E T}.
\end{multline} 
where $\bar{n}=\sqrt{6(n-4)/(n-5)}$ and $\lambda$, $k$ and $K$ are constant. Here we take $E>k+\lambda$.
By superposing the function $\psi_{\lambda,E,k}(u,v,\phi,T)$, the wavepacket can be formed as;
\begin{multline}\label{wp}
\Psi(u,v,\phi,T)=\int_0^\infty d\lambda \int_0^\infty dk \int_0^\infty dE'\sin(u\sqrt{\frac{\lambda}{2-\bar{n}}})\\ \times\sin(v\sqrt{\frac{E'}{2+\bar{n}}})\sin(\phi\sqrt{4\omega k})e^{- (\lambda+E'+k) (\gamma+iT)},
\end{multline}
where $E'=E-k-\lambda$ and $\gamma$ is a positive constant. \\Norm $\| \Psi\|$ can be calculated as,
\begin{equation}
\| \Psi\|=\int \Psi \Psi^* d\bar{u}d\bar{v}d\phi=\frac{1}{32 \sqrt{2\omega}}\left(\frac{\pi}{\gamma}\right) ^{9/2}.
\end{equation}
This clearly shows that we have a finite and time-independent norm.
From equation (\ref{wp}), the normalized wavepacket can be written as,
\begin{multline}
\Psi(\bar{u},\bar{v},\phi,T)=\biggl( \frac{2\omega}{\pi} \biggr)^{3/4} \biggl( \frac{\sqrt{\gamma} }{\gamma+iT} \biggr)^{9/2} \\ \times \bar{u} \bar{v} \phi exp \biggl[-\frac{1}{4}\biggl(\frac{4 \omega \phi^2+\bar{u}^2+\bar{v}^2}{ \gamma+i T}\biggr)\biggr],
\end{multline}
where $\bar{u}=\frac{u}{\sqrt{|2-\sqrt{6(n-4)/(n-5)}|}}$ and $\bar{v}=\frac{v}{\sqrt{2+\sqrt{6(n-4)/(n-5)}}}$.\\
Now we can calculate the expectation values of $a$, $b$ and the proper volume measure $V= (-g)^{\frac{1}{2}}=a^3b^{n-4}$ in terms of $\bar{u}$ and $\bar{v}$ as,\\
\begin{multline}\label{ea}
\langle a \rangle =\int_{-\infty}^\infty d\bar{u}\int_{-\infty}^\infty d\bar{v} \int_{-\infty}^\infty d\phi |\Psi(\bar{u},\bar{v},\phi,T)|^2 \\ \times exp\biggl[\frac{1}{2}\sqrt{\frac{n-5}{n-2}}\biggl(\sqrt{\left|2-\sqrt{\frac{6(n-4)}{(n-5)}}\right|}\bar{u}\\ +\sqrt{2+\sqrt{\frac{6(n-4)}{(n-5)}}}\bar{v}\biggr)\biggr] ,
\end{multline}
\begin{multline}\label{eb}
\langle b \rangle =\int_{-\infty}^\infty d\bar{u}\int_{-\infty}^\infty d\bar{v} \int_{-\infty}^\infty d\phi |\Psi(\bar{u},\bar{v},\phi,T)|^2 \\ \times exp \biggl[\frac{1}{2}\sqrt{\frac{1}{(n-2)(n-4)}}\biggl(\sqrt{\left|2-\sqrt{\frac{6(n-4)}{(n-5)}}\right|}\bar{u} \\ -\sqrt{2+\sqrt{\frac{6(n-4)}{(n-5)}}}\bar{v}\biggr)\biggr],
\end{multline}
\begin{multline}\label{ev}
\langle V \rangle =\int_{-\infty}^\infty d\bar{u}\int_{-\infty}^\infty d\bar{v} \int_{-\infty}^\infty d\phi |\Psi(\bar{u},\bar{v},\phi,T)|^2 \\ a^3(\bar{u},\bar{v})b^{n-4}(\bar{u},\bar{v}). 
\end{multline}
Lengthy expressions of the above integrals have not been mentioned here. Therefore, plots for the same are shown in the figures taking the value of positive constant $\gamma=1$.  
\begin{figure}[hbtp]
\centering
\includegraphics[scale=0.6]{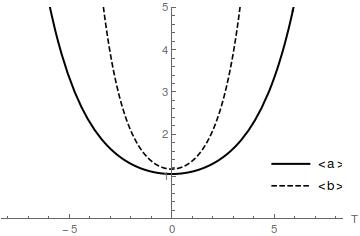}
\caption{Expectation value of scale factors for n=6.}
\end{figure}
\begin{figure}[hbtp]
\centering
\includegraphics[scale=0.6]{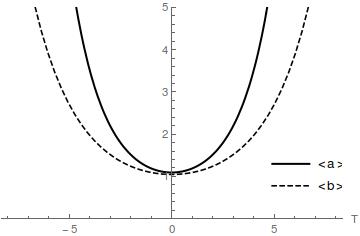}
\caption{Expectation value of scale factors for n=8 .}
\end{figure}
\begin{figure}[hbtp]
\centering
\includegraphics[scale=0.6]{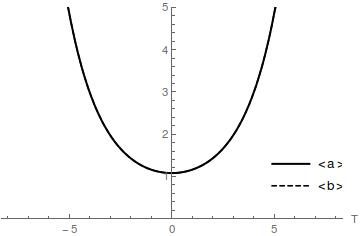}
\caption{Expectation value of scale factors for n=7 is given above and as it is shown, they both coincide.}
\end{figure}
\begin{figure}[hbtp]
\centering
\includegraphics[scale=0.6]{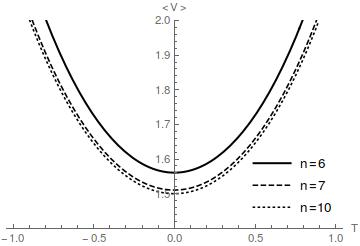}
\caption{Expectation value of Volume $<V=a^3b^{n-4}>$ .}
\end{figure}

The figures clearly indicate that there is no singularity of a zero proper volume. In fact both the scale factors $a$ and $b$ are quite well behaved individually.
\subsection{For stiff  fluid $\alpha=1$ and $n=5$}
For $n=5$ with $\alpha=1$,equation(\ref{wd}) will become 
\begin{multline} \label{wdw3}
\biggl[\frac{1}{6}\frac{\partial^2}{\partial A \partial B }-\frac{1}{6}\frac{\partial^2}{\partial B^2}-\frac{1}{4\omega}\frac{\partial^2}{\partial \phi^2}-i\frac{\partial }{\partial T} \biggr] \psi(A,B,\phi,T)=0.
\end{multline}
With the change of variable as $x=2A+B$ and $y=B$, it can be re-written as,
\begin{multline} \label{wdw2}
\biggl[-\frac{1}{6}\frac{\partial^2}{\partial x^2}+\frac{1}{6}\frac{\partial^2}{\partial y^2}+\frac{1}{4\omega}\frac{\partial^2}{\partial \phi^2}+ E \biggr] \xi(x,y,\phi)=0.
\end{multline}
Solution of above equation can be found using similar method what was done in the previous section. Normalized wavepacket is given as,
\begin{multline}  \label{wp1}
\Psi(\bar{x},\bar{y},\phi,T)=\biggl( \frac{2\omega}{\pi} \biggr)^{3/4} \biggl( \frac{\gamma^{3/2} }{(\gamma^2+T^2)(\gamma+iT)} \biggr)^{3/2} \\ \times \bar{x} \bar{y} \phi exp \biggl[-\frac{1}{4}\biggl(\frac{\bar{x}^2}{ \gamma-i T}+\frac{4 \omega \phi^2+\bar{y}^2}{ \gamma+i T}\biggr)\biggr],
\end{multline}
where $\bar{x}=\sqrt{6}x$ and $\bar{y}=\sqrt{6}y$.\\
The relevant expectation values are given as 
\begin{eqnarray} \label{eab}
\langle a \rangle= \frac{e^{\frac{\gamma^2+T^2}{24 \gamma}} \left(\gamma (\gamma+24)+T^2\right)^2}{576 \gamma^2},\\\label{eabc}
\langle b \rangle=\frac{e^{\frac{\gamma^2+T^2}{12 \gamma}} \left(\gamma (\gamma+6)+T^2\right)}{6 \gamma}.
\end{eqnarray}
\begin{figure}[hbtp]
\centering
\includegraphics[scale=0.6]{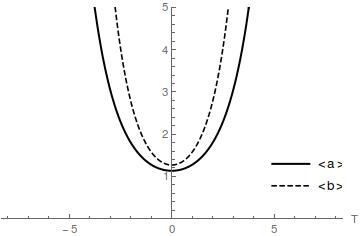}
\caption{Expectation value of scale factors for n=5.}
\end{figure}
\subsection{For fluid with $\alpha\neq 1$}
We will be considering  only $n=7$ as an example for a general fluid with $\alpha\neq 1$, in which case the equation is tractable. Now Wheeler-DeWitt equation (\ref{wdw}) can be written as,
\begin{multline} \label{wdw2}
\biggl[-\frac{\partial^2}{\partial u^2}+5\frac{\partial^2}{\partial v^2}+\frac{1}{4\omega}\frac{\partial^2}{\partial \phi^2}+e^{\frac{3\alpha-3}{\sqrt{30}}u} E \biggr] \xi(u,v,\phi)=0.
\end{multline}
With suitable operator ordering as  done in work by Pal and Banerjee\cite{sp1,sp2}, equation (\ref{wdw2}) takes following form, 
\begin{multline}
\biggl[ e^{\frac{(3-3\alpha)}{2\sqrt{30}}u}\frac{\partial}{ \partial u}e^{\frac{(3-3\alpha)}{2\sqrt{30}}u}\frac{\partial}{ \partial u}  -5e^{\frac{(3-3\alpha)}{\sqrt{30}}u}\frac{\partial^2}{\partial v^2}\\ -e^{\frac{(3-3\alpha)}{\sqrt{30}}u} \frac{1}{4\omega}\frac{\partial^2}{\partial \phi^2}-E \biggl] \xi(u,v,\phi)=0.
\end{multline} 
Now with ansatz $\xi(u,v,\phi)=U(u)e^{iV_0v/\sqrt{5}}e^{i\sqrt{4\omega}\phi}$, above equation can be re-written as,
\begin{multline}
\biggl[ e^{\frac{(3-3\alpha)}{2\sqrt{30}}u}\frac{\partial}{ \partial u}e^{\frac{(3-3\alpha)}{2\sqrt{30}}u}\frac{\partial}{ \partial u} +e^{\frac{(3-3\alpha)}{\sqrt{30}}u}(V_0^2+k^2) \\-E \biggl] U(u)=0.
\end{multline}
With change of variable like $\chi= e^{\frac{(3\alpha-3)}{2\sqrt{30}}u}$, this equation can be written as,
\begin{multline} \label{henkel}
\biggl[ \frac{3(\alpha-1)^2}{40}\frac{\partial^2}{ \partial \chi^2}+\frac{V_0^2+k^2}{\chi^2} -E \biggr] U(\chi)=0 .
\end{multline}
Above equation is similar to the inverse square potential problem of physics which admits the self-adjoint extensions. The solution to equation(\ref{henkel}) can be given by Hankel functions. It has been extensively shown in the work\cite{sp1}.

\section{Discussion with Concluding Remarks}

In this work, using the n-dimensional generalization of Schutz formalism\cite{ndfrw}, we formulated the Lagrangian and Hamiltonian for our model. Followed by multiple canonical transformations, we obtained the complicated looking  Wheeler-DeWitt equation(\ref{wdw}) for the wavefunction of the n-dimensional anisotropic universe. We obtained the general wavepacket of Wheeler-DeWitt equation for stiff fluid $\alpha=1$.  We are able to find a time independent and finite normed wavepacket which establishes the unitary evolution of the model.

We have found the non-singular expectation values of the scale factors and these are shown in Fig. 1, 2, 3 and 5 for different n.  Expectation values of scale factors $a$ and $b$ coincide for $n=7$.  This is not surprising as both usual space-section and the extra 3-dimensional space, although anisotropic between each other to start with, are 3-dimensional isotropic space like sections in themselves.
 
 Non-zero minima of the expectation values of volume element(Fig. 4) clearly show the contraction of the universe followed by expansion avoiding any singularity indicating a bouncing universe. Similar results were obtained in quantization of anisotropic models in absence of fluid in the past \cite{ndaniso}. 

For a general fluid with $\alpha \neq 1$, we are able to show the self-adjoint extension and thus unitary evolution of the model for $n=7$ despite the mathematical difficulty of the model. However, the result of it is in accordance with four-dimensional anisotropic model Bianchi-I as expected \cite{sp1}.

Our results may also be considered as the higher dimensional generalization of unitary evolution of the anisotropic models as shown in \cite{sp1,sp2,sp3,sachin}.\\

\section*{Acknowledgements} 
The author would like to thank Narayan Banerjee for insightful discussions and comments in preparing the manuscript.
The author was financially supported by CSIR, India.  \\



\begin{thebibliography}{99}
\bibitem{kalu} T. Kaluza, Sitzungsber. Preuss. Akad. Wiss. Phys. Math. Kl. {\bf 33 }, 966 (1921).
\bibitem{klein} O. Z. Klein, Physik {\bf 37}, 895 (1926).
\bibitem{klein1} O. Z. Klein, Nature {\bf 118}, 516 (1926).
\bibitem{o1} Y. M. Cho, J. Math. Phys. (N.Y.) {\bf 16}, 2029 (1975).
\bibitem{o2} Y. M. Cho and P. G. O. Freund, Phys. Rev. D {\bf 12}, 1711 (1975).
\bibitem{ss}M. B. Green, J. H. Schwarz, and J. H. Witten, Superstring Theory(Camb. Univ Press, Cambridge, 1987).
\bibitem{sg} I. L. Buchbinder and S. M. Kuzenzo, Ideas and Methods of Supersymmetry and Supergravity(IOP Publ., Bristol and Philadelphia, 1995).
\bibitem{witt} B S DeWitt, Phys. Rev. {\bf 160}, 1113 (1967). 
\bibitem{wheeler} J A  Wheeler, Superspace and the nature of quantum geometrodynamics Batelle Recontres (New York: Benjamin, 1968). 
\bibitem{misner} C W Misner, Phys. Rev. {\bf 186}, 1319 1969.

\bibitem{wilt} D L Wiltshire,   \href{arXiv:0101003 [gr-qc]}{arXiv:0101003}. 
\bibitem{jj} J J Halliwell, Quantum Cosmology and Baby Universes ed S Coleman, J B Hartle, T Piran and S Weinberg (Singapore: World Scientific,1991).
\bibitem{pinto} N Pinto-Neto  and J C Fabris   Class. Quant. Grav. {\bf30}, 143001 (2013).
\bibitem{nu1}J E Lidsey,  Phys. Lett. B {\bf 352}, 207 (1995).
\bibitem{nu2}N Pinto-Neto, A F Velasco  and Jr R Collistete,   Phys. Lett. A {\bf 277}, 194 (2000).
\bibitem{nu3}K V Kuchar,  Conceptual Problems in Quantum Gravity ed A Ashtekar and J Stachel (Boston: Birkhause, 1991).
\bibitem{nu4} C J Isham, Integrable Systems, Quantum Groups and Quantum Field Theory ed L A Ibort and M A Rodriguez (Dordrecht: Kluwer, 1993).
\bibitem{nu5}C Rovelli, Found. Phys.  {\bf 41}, 1475 (2011);\href{arXiv:0903.3832 [gr-qc]}{arXiv:0903.3832}.
\bibitem{nu6} E Anderson, \href{arXiv:1009.2157 [gr-qc]}{arXiv:1009.2157}.
\bibitem{schutz1} B. F. Schutz, Phys. Rev. D {\bf 2}, 2762 (1970).
\bibitem{schutz2} B. F. Schutz, Phys. Rev. D {\bf 4}, 3559 (1971).
\bibitem{is1} F. G. Alvarenga and N. A. Lemos, Gen. Relativ. Gravit.{\bf 30}, 681 (1998).
\bibitem{is2} A. B. Batista, J. C. Fabris, S. V. B. Goncalves, and J. Tossa,Phys. Rev. D {\bf 65}, 063519 (2002).
\bibitem{is3} F. G. Alvarenga, J. C. Fabris, N. A. Lemos, and G. A.Monerat,Gen. Relativ. Gravit. {\bf34}, 651 (2002).
\bibitem{vakili1} B. Vakili  Physics Letters B {\bf 688}, 129136  (2010).
\bibitem{sp1} S. Pal and N. Banerjee, Phys. Rev. D, {\bf 90}, 104001 (2014); \href{arXiv:1410.2718v1 [gr-qc]}{arXiv:1410.2718v1}.
\bibitem{sp2} S. Pal and N. Banerjee, Phys. Rev. D {\bf 91}, 044042  (2015); \href{arXiv:1411.1167v2 [gr-qc]}{arXiv:1411.1167v2}.
\bibitem{sp3} S. Pal, Class. Quant. Grav. {\bf 33}, 045007  (2016); \href{arXiv:1504.02912 [gr-qc]}{arXiv:1504.02912}.
\bibitem{sachin} S. Pandey and N. Banerjee, Phys. Scr. {\bf 91 } 115001 (2016); \href{arXiv:1603.01406 [gr-qc]}{arXiv:1603.01406}.
\bibitem{sp5} S. Pal and N. Banerjee, J. Math. Phys. {\bf 57}, 122502  (2016); \href{arXiv:1601.00460 [gr-qc]}{arXiv:1601.00460}.
\bibitem{brans} C. Brans and R.H. Dicke, Phys. Rev., {\bf 124}, 925 (1961).
\bibitem{dicke} R.H. Dicke, Phys. Rev., {\bf 125}, 2163 (1962).

\bibitem{vakili2} B. Vakili, Phys. Lett. B, {\bf 718}, 34 (2012).
\bibitem{sachin1}  S. Pandey and N. Banerjee; Eur. Phys. J. Plus {\bf 132}, 107 (2017).
\bibitem{sachin2}  S. Pandey, S. Pal and N. Banerjee; Annals Phys. {\bf 393}  93-106 (2018).
\bibitem{bdfluid} C. R. Almeida, A. B. Batista, J. C. Fabris and P. V. Moniz, J. Math. Phys. {\bf 58}, 042301 (2017).
\bibitem{sp4} S. Pal, Phys. Rev. D {\bf 94}, 084023 (2016).
\bibitem{ndaniso} F.A.P. Alves-Junior, M.L. Pucheu, A.â€‰B. Barreto, and C. Romero, Phys. Rev. D {\bf 97}, 044007 (2018).
\bibitem{ndfrw} P. S. Letelier and J. P. M. Pitelli,  Phys. Rev. D {\bf 82}, 104046 (2010).
\bibitem{Lapchinskii} V. G. Lapchinskii and V. A. Rubakov, Theor. Math. Phys. {\bf 33}, 1076 (1977).
\bibitem{gold} H. Goldstein, Classical Mechanics, (Addison-Wesley Pub. Co., USA, 1959)
\end{thebibliography}
\end{document}